\documentclass[%
  prb,%
  letterpaper,%
  twocolumn,%
  showpacs,%
  floatfix,%
  superscriptaddress%
]{revtex4}

\usepackage{graphicx}
\usepackage{dcolumn}
\usepackage{bm}
\usepackage{amsmath}

\begin{document}

\preprint{APS/123-QED}

\title{A critical analysis on the sensitivity enhancement of surface plasmon resonance sensors with graphene}

\author{Aline dos S. Almeida}
\author{D. A. Bahamon}
\affiliation {School of Engineering, Mackenzie Presbyterian University, S\~ao Paulo - 01302-907, Brazil}
\affiliation {MackGraphe, Mackenzie Presbyterian Institute, S\~ao Paulo - 01302-907, Brazil}
\author{Nuno M. R. Peres}%
\affiliation{Physics Department - Minho University, Gualtar’s Campus - 4710-057 Braga, Portugal}
\author{Christiano J. S. de Matos}
\email{cjsdematos@mackenzie.br}
\affiliation {School of Engineering, Mackenzie Presbyterian University, S\~ao Paulo - 01302-907, Brazil}
\affiliation {MackGraphe, Mackenzie Presbyterian Institute, S\~ao Paulo - 01302-907, Brazil}

\begin{abstract}
The use of graphene in surface plasmon resonance sensors, covering a metallic (plasmonic) film, has a number of demonstrated advantages, such protecting the film against corrosion/oxidation and facilitating the introduction of functional groups for selective sensing. Recently, a number of works have claimed that few-layer graphene can also increase the sensitivity of the sensor. However, graphene was treated as an isotropic thin film, with an out-of-plane refractive index that is identical to the in-plane index. Here, we critically  examine the role of single and few layers of graphene in the sensitivity enhancement of surface plasmon resonance sensors. Graphene is introduced over the metallic film via three different descriptions: as an atomic-thick two-dimensional sheet, as a thin effective isotropic material (same conductivity in the three coordinate directions), and as an non-isotropic layer (different conductivity in the perpendicular direction to the two-dimensional plane). We find that only the isotropic layer model, which is known to be incorrect for the optically modelling of graphene, provides sizeable sensitivity increases, while the other, more accurate, models lead to negligible contribution to the sensitivity.
\end{abstract}

\maketitle


\section{\label{sec:level1}Introduction}

Lately, much attention has been paid to the development of new chemical and biological sensors based on low cost and fast diagnosis\cite{BiosensorReview1:21,BiosensorReview2:20,BiosensorSpeci1:21,BiosensorSpeci2:21}. Surface plasmon resonance (SPR) is a high sensitivity, real time, and label free technique with great potential for sensing applications\cite{Pathak:19}. Surface plasmon polaritons (SPPs) are surface waves propagating at a metal/dielectric interface and arise from the electromagnetic field coupling to dipolar excitation of free electrons in the  metal. In the resonance condition, these waves are highly sensitive to changes in dielectric function the surrounding environment \cite{Pathak:19,Maier}. In particular, for SPR biosensors, reversible adsorption  of biomolecules (analyte) onto the sensing surface (metal/dielectric interface) changes the local refractive index (RI), leading to changes in the SPR condition  \cite{Tang:10}.

A conventional SPR biosensor consists of a thin metallic film deposited on one side of a prism. The metallic film is typically made up from gold, which supports the propagation of SPPs at the visible wavelength range and has good resistance to oxidation and corrosion in wet environments. However, sensing specificity usually requires preparing the surface (e.g., using alkanethiol in the case of gold \cite{Choi:10}) to receive a linker molecule which, in turn, allows for the fixation of the required bio-receptor \cite{Choi:10,Mok:08,Taninaka:10}. In this sense, adding a graphene layer to the metal surface may simplify functionalization steps, as graphene can work as a bio-functional surface due to ${\pi}-{\pi}$ interactions with carbon-based ring structures that are widely present in bio-molecules \cite{Song:10}. Besides, it has already been experimentally demonstrated that graphene can protect the surface against oxidation and corrosion, which is particularly important when more reactive metals, such as copper, a cheaper alternative to gold, is used as the plasmonic material \cite{GRIGORENKO:14,GRIGORENKO:19}.

In addition to the mentioned advantages, recently there have been theoretical reports that seem to indicate that graphene layers could improve gold's \cite{Chen:19} and silver's \cite{PAL:21,Choi:11} SPR sensitivity. Nevertheless calculated sensitivity increases diverge significantly from report to report, ranging from 1 to 200${\%}$ for few-layer graphene \cite{Bhavsar:19,Wu:10,SAIFUR:17}. We note that these discrepancies are rather intriguing, as the same physical parameters and SPR system geometry seem to have been used, with only one report presenting a slight difference: a thin binding layer between graphene and the analyte, producing some absorption.\cite{Wu:10} We also highlight that all three reports treat graphene as an ultrathin isotropic material, which is notorious for yielding un-physical electromagnetic results \cite{Matthes:16,Majerus:18,Valuev16,Oliveira:15}. According to Refs. \cite{Valuev16,Majerus:18}, modeling graphene as an isotropic thin film results in an artificial plasmonic resonance (of a metallic nature), since the same refractive index is adopted in and out of the plane where graphene is placed. In this case, the electric field of light in the \emph{p}-polarization, which has a component that is normal to the graphene plane, generates a non-physical current inside the graphene thin film along the perpendicular direction.

In this work, to access the actual impact of graphene on the SPR sensor sensitivity. We model graphene in a number of ways and compare the obtained results. First, graphene is modeled as a surface conductivity, in which case it is treated as a strictly 2D material. Subsequently, graphene is treated as an ultrathin isotropic film, and, finally, as an ultrathin anisotropic  film.  Our results show that the selected modelling approach directly impacts on the observed sensitivity enhancement, with significant enhancements obtained only in the isotropic case. Furthermore, in the case of an isotropic ultrathin layer, the results highlight the importance of the choice of the value for graphene's refractive index, with results diverging more than $\sim 50\%$ among them.

\begin{figure}
\scalebox{1.2}{\includegraphics{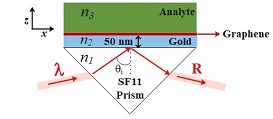}}
\caption{Schematic representation of the SPR sensor.}
\label{fig:prisma}
\end{figure}

\section{\label{sec:level2}Model Description}

The proposed SPR sensor structure is identical to that examined in Refs.\cite{Bhavsar:19,SAIFUR:17}, and is based on the Kretschmann configuration. As sketched in Fig. \ref{fig:prisma}, it consists of a multilayer structure: a SF11 glass prism, a 50-nm-thick gold film, a graphene layer, and the analyte. In this configuration, the SPR resonance appears as a dip in the reflectance of \emph{p}-polarized light impinging on the prism at a certain angle larger than the critical angle of the prism-analyte interface. Thus, the SPR condition is easily calculated through the reflectance of the multilayer structure ($R = |r_{1...N}|^2$). Note that we treat the structure as a three-layer system when graphene is modelled as a surface conductivity, and as a four-layer system when graphene is taken to be a thin film.

\begin{figure}
\scalebox{1}{\includegraphics{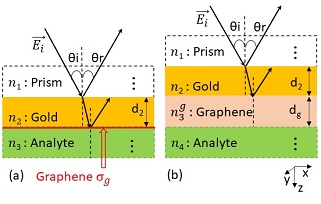}}
\caption{(a) Three-layer system: graphene as an atomic sheet with a surface conductivity $\sigma_g$. (b) Four-layer system: graphene as a thin layer film with thickness d$_g$ and refractive index $n^g_3$.}
\label{fig:m2D3D}
\end{figure}

For the three-layer system (2D graphene model; see Fig.\ref{fig:m2D3D}(a)), the first layer is the SF11 glass prism (semi-infinite) with refractive index ${n_{1}}$ = 1.7786  coated by a gold film with refractive index ${n_{2}}$  = 0.1834 + 3.4332\emph{i} \cite{Bhavsar:19} and the third layer is the aqueous analyte that is considered  semi-infinite (water refractive index taken to be 1.332). In the four-layer case (3D graphene model presented in Fig. \ref{fig:m2D3D}(b)),  graphene is considered as the third layer with thickness $d_g$ and refractive index $n^g_3$, while the semi-infinite  region of the analyte is now considered as the fourth layer, with refractive index $n_4$. The $n^g_3$ parameter will be discussed in detail below, for the isotropic and anisotropic cases. The reflection coefficients for the three- and four-layer systems are respectively given by:
\begin{eqnarray}
r_{123}=\ \frac{r_{{12}\ }\ +\ r_{{23}}\ e^{i2\phi_2}\ }{{1\ +\ r}_{{12}}\ r_{{23}}\ \ e^{i2\phi_2}}
\label{eq:r13}
\end{eqnarray}
and 
\begin{eqnarray}
r_{{1234}}=\frac{\left(r_{{12}}+r_{{23}}e^{i2\phi_2}\right)\ +\left(r_{12}r_{23}+e^{i2\phi_2}\right)r_{{34}}e^{i2\phi_3}}{\left({1+r}_{{12}}\ r_{{23}}\ e^{i2\phi_2}\right)+\left(r_{{23}}+r_{{12}\ }e^{i2\phi_2}\right)r_{{34}}e^{i2\phi_3}},
\label{eq:r14}
\end{eqnarray} 
where $r_{ij}$ are the Fresnel reflection coefficients for \emph{p}-polarized  waves between layers $i$ and $j$, $\phi_i = k_{zi}d_i$ and $k_{zi} = \sqrt{n^2_i\frac{\omega^2}{c^2} - k^2_{xi}}$. Given that the layered systems present translational symmetry along the  \emph{x}-direction, $k_{xi} = k_x = n_1\left(\frac{\omega}{c} \right )\sin \theta_i$. $\omega$ and $c$ are respectively the angular frequency and the speed of light in vacuum.

The angle corresponding to the reflectance minimum is called SPR angle $\left(\theta_{SPR}\right)$. The sensing mechanism consists of monitoring the $\theta_{SPR}$ shift as the refractive index of the analyte (sensing medium) changes \cite{WIJAYA:11}. We  assume that the refractive index of the aqueous analyte $n_{3(4)}$ varies  from 1.332 to 1.35 in steps of ${\Delta n}={5\times10^{-6}}$. The sensor's sensitivity is defined as $S =\frac{d\theta_{SPR}}{dn_{3(4)}}$, in units of $^{\circ}$/RIU (where RIU stands for Refractive Index Units), and is calculated for each studied case. To clearly access the effect of the graphene modelling on the performance of the sensitivity, the thickness of the gold layer ($d_2$) and the wavelength of the incident light ($\lambda = 633$ nm) are kept constant.

\subsection{\label{sec:level3}Graphene Modeling}

In the three-layer case, graphene is modelled as an atomically thin film. Its surface conductivity ($\sigma_g$) is introduced in the boundary conditions of the electromagnetic fields at the metal-analyte interface to obtain the Fresnel reflection coefficient 
\begin{eqnarray}
r_{23}=\frac{k_{z_2}\epsilon_3\ -\  k_{z_3}\epsilon_2\ +\ \frac{k_{z_2}k_{z_3}\sigma_{g}}{\epsilon_0\omega}}{k_{z_2}\epsilon_3\ +\ k_{z_3}\epsilon_2\ +\ \frac{k_{z_2}k_{z_3}\sigma_{g}}{\epsilon_0\omega}}.
\label{eq:r23g}
\end{eqnarray} 
In this expression, $\epsilon_0$ is the vacuum permittivity. In the visible region of the electromagnetic spectrum  ${\sigma_{g}\approx \sigma}_0\ =\ \frac{e^2}{4\hbar}$ \cite{Nair:08,Peres:15}, with  $e$ being the electron charge and $\hbar$ the reduced Planck constant.

In the isotropic 3D model, graphene is considered a thin film with an effective thickness $d_g = 0.34$ nm \cite{Bhavsar:19,Wu:10,SAIFUR:17,Oliveira:15,Ni:07} and refractive index
\begin{eqnarray}
n^g_3=\sqrt{1+i\frac{\sigma_{g}}{\epsilon_0\omega d_g}}=2 + 1.7119i.
\label{eq:ng3diso}
\end{eqnarray}
Finally, in the anisotropic case, graphene is taken to have different in-plane and out-of-plane values for the diagonal dielectric tensor components. Specifically, for the in-plane response, the refractive index was considered to have the same value as in the isotropic case $ n^g_{3\text{in}} = n_{xx} = n_{yy}=\sqrt{\epsilon^g_{3\text{in}}}=\sqrt{1+i\frac{\sigma_{g}}{\epsilon_0\omega d_g}}$. The out-of-plane refractive index was assumed to be $n^g_{3z} = \sqrt{\epsilon^g_{3z}} = \sqrt{2.5}$. \cite{Oliveira:15,Gao:12,Kwon:14} Considering that the reflecting surface of the anisotropic thin film coincides with graphene's basal plane, the  coupling between the \emph{s}- and \emph{p}-polarized waves is zero. This ensures that the reflectance of the four-layer system with the anisotropic graphene continues to be calculated by Eq. \ref{eq:r14}, but with $\phi_3 = k_{z3e}d_3$,

\begin{eqnarray}
k_{z3e}=\sqrt{\left(\epsilon^g_{3\text{in}}\frac{\omega^2}{c^2}-\frac{\epsilon^g_{3\text{in}}}{\epsilon^g_{3z}} k_x^2\right)},
\label{eq:kz3e}
\end{eqnarray}

\noindent and

\begin{eqnarray}
r_{i3} = \frac{\left(n^g_{3\text{in}}\right)^2k_{zi}-n^2_ik_{z3e}}{\left(n^g_{3\text{in}}\right)^2k_{zi} + n^2_ik_{z3e}}.
\label{eq:ri3ani}
\end{eqnarray}

\section{\label{sec:level4}Results and Discussion}

\begin{figure}
\scalebox{0.8}{\includegraphics{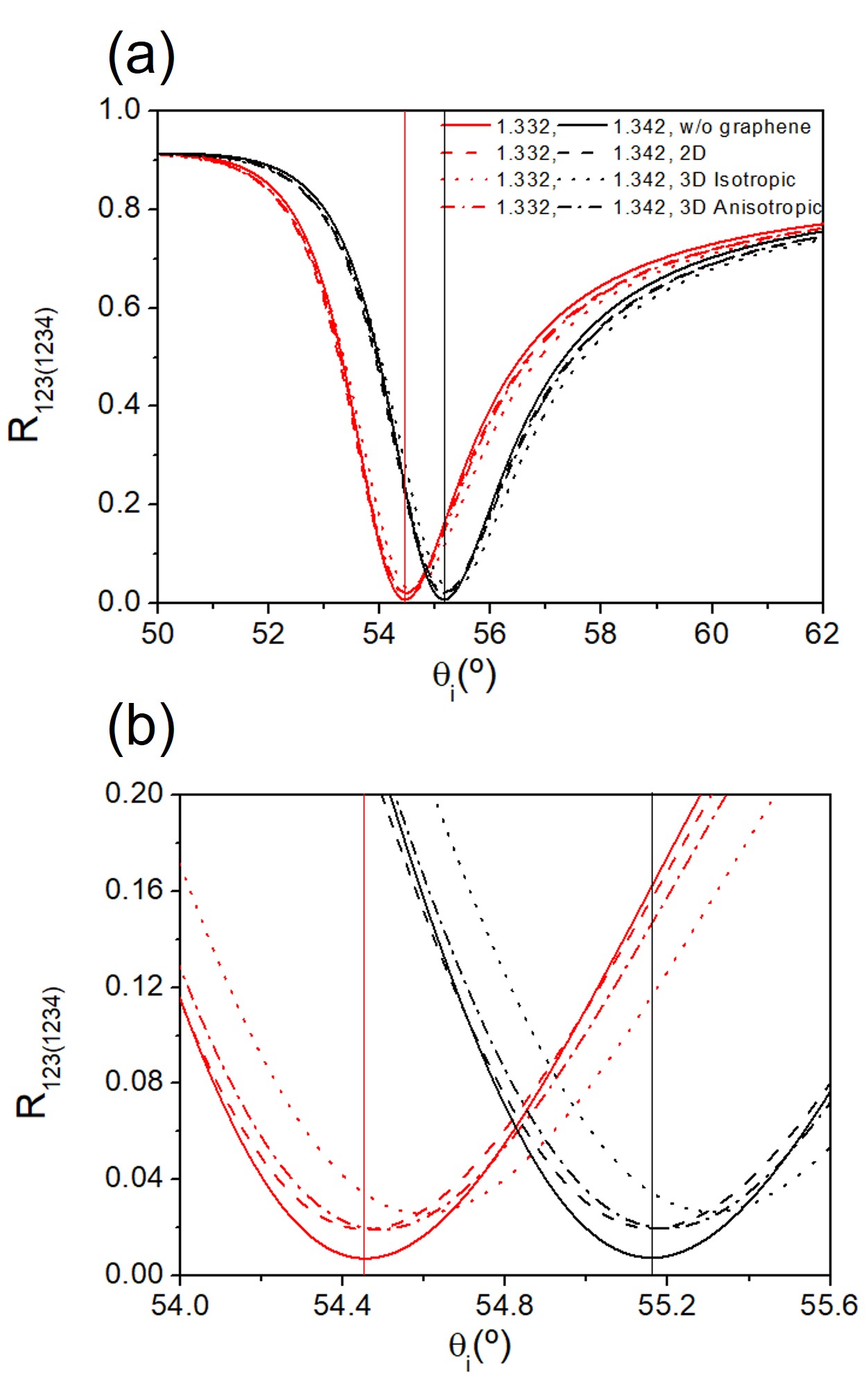}}
\caption{\label{fig:RpMono} (a) Calculated reflectance $(\text{R}_{123(1234)})$ as a function of the incident angle $(\theta_i)$ for analyte refractive indices of 1.332 and 1.342 (red and black curves, respectively). The vertical lines indicate the positions of $(\theta_{SPR})$ for the system without graphene. Graphene modelled as a surface conductivity (dashed lines); and as isotropic $(n^g_{3})$ (dotted lines) and anisotropic $(n^g_{3\text{in}})$ (dashed-dotted lines) films. (b) Zoom in the the minimum reflectance region of (a).}
\label{fig:Rmono}
\end{figure}

The effect of the different approaches used to model the graphene monolayer is presented in Fig.\ref{fig:Rmono}(a), which shows the reflectance for the three-layer ($R_{123}$) and the four-layer ($R_{1234}$) systems as functions of the angle of incidence, $\theta_i$, for two different analyte refractive indices: $n_{3(4)} = 1.332~\text{and}~1.342$ (red and black curves, respectively). The continuous lines correspond to the response of the sensor without graphene; the dashed lines present the response with graphene as a sheet conductivity (2D model); the dotted lines stand for the 3D isotropic case; and the dashed-dotted lines represent the 3D anisotropic model. 

At first sight, the four models produce similar reflectance curves, with graphene only slightly modifying the response of the sensor. However, the zoom in the reflectance curves provided in Fig. \ref{fig:Rmono}(b) shows observable differences produced by the different graphene models, particularly in the shift of $\theta_{SPR}$ with the analyte refractive index, as well as in the amplitude of the reflectance at the plasmon resonance condition. For example, in the system based only on gold's SPR, the SPR corresponds to a minimum reflectance of $\sim 0.71\%$ for the analyte refractive index of 1.332 (solid red curve in Fig. \ref{fig:Rmono}). By adding graphene monolayer and modeling it either through the 2D model or the anisotropic thin film model, the reflectance minimum increases to $\sim 1.9\%$ due to the absorption by the sheet of graphene, while modelling graphene as a thin isotropic layer produces not only the largest shift of $\theta_{SPR}$, but also the largest change in the reflectance minimum, which increases to $\sim 2.6\%$. 

However, the most important result of the present work is to quantify the sensitivity of the sensor (i.e., the SPR shift with analyte index change), which is shown in table \ref{table1} for the different graphene models ($S_{gr}$), as well as for the sensor without graphene ($S_{Au}$). The sensitivity increase provided by graphene, given by $\Delta S = \left((S_{gr}-S_{Au})/S_{Au}\right)\times 100$, is also presented for each case. When modelling monolayer graphene as an atomic sheet (2D model) or as an anisotropic thin film (3D anisotropic) there is minimal change in the system's sensitivity relative to case of gold only. In stark contrast, modelling monolayer graphene as an isotropic film, which is known to lead to unrealistic results, as discussed \cite{Oliveira:15}, yields a sensitivity increase that is 3 orders of magnitude higher than with the 2D model. 

\begin{table}
\caption{\label{table1}%
Sensitivity and sensitivity increase for the SPR sensor without and with a monolayer of graphene modeled in the three different examined ways.}
\begin{ruledtabular}
\begin{tabular}{ccc}
Model&$S_{Au(gr)}(^{\circ}/\text{RIU}$)& $\Delta S$ (\%) \\
\colrule
Gold & 70.8951 & 0\\
2D & 70.8962 & 0.0016\\
3D Iso. &  71.4515 & 0.7848\\
3D Aniso. & 70.8339 & -0.086
\end{tabular}
\end{ruledtabular}
\end{table}

\begin{figure}
\scalebox{0.85}{\includegraphics{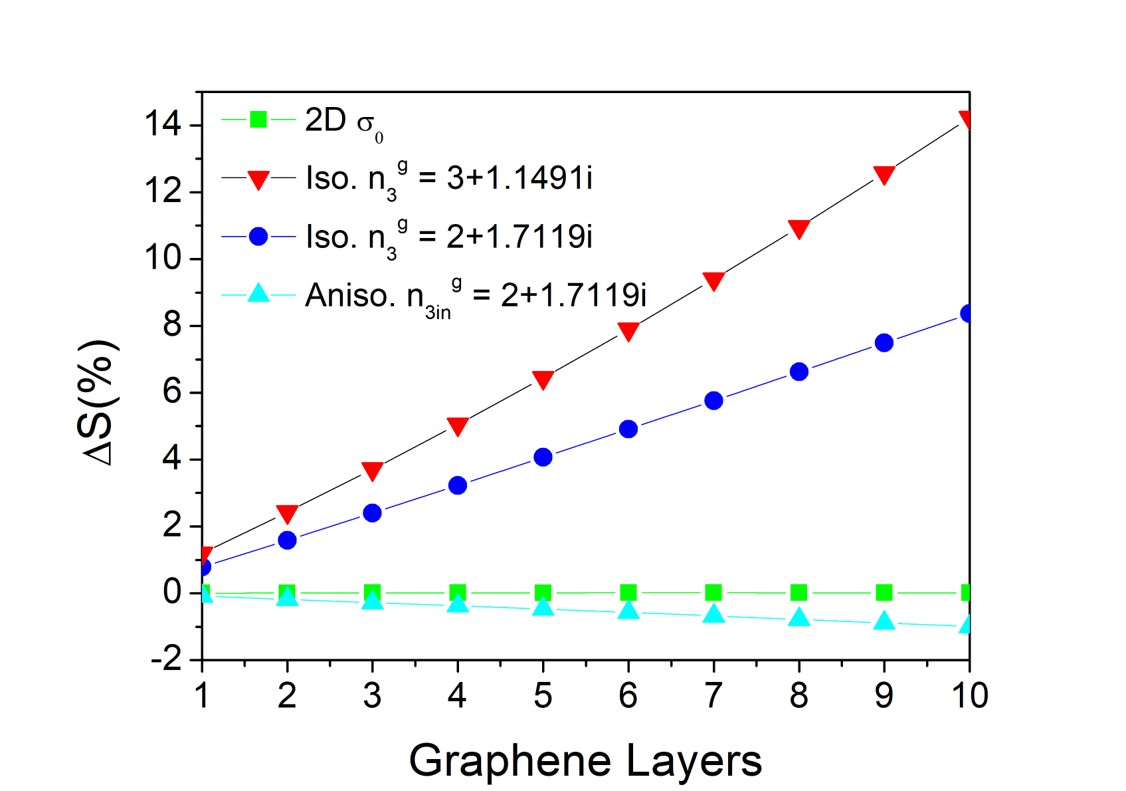}}
\caption{Calculated sensitivity increase for the addition of graphene layers compared with the system based only on gold. Graphene as an: atomic sheet (green), isotropic film, $n^g_{3}  = 3+1.1491i$ (dark blue) and $n^g_{3}= 2 + 1.7119i$ (red); anisotropic film,  $n^g_{3\text{in}}= 2 + 1.7119i$ (cyan).}
\label{fig:Sensi} 
\end{figure}

Having established that the way a single sheet of graphene is modelled affects the obtained (theoretical) sensor sensitivity, we now investigate the sensitivity as a function of the number of graphene layers. To model more than one layer, previous works treating graphene as a thin film \cite{PAL:21,Choi:11,Bhavsar:19,Wu:10,SAIFUR:17} have used the transfer matrix method, with each graphene layer represented by a individual matrix. This strategy is  equivalent to the one we proposed using eq. \ref{eq:r14}, given that the multiplication of transfer matrices of the N-layers of graphene ultimately leads to an  effective layer of thickness $L = N\times d_g = N\times0.34$ nm, with the same refractive index $n^g_3$ of the monolayer. In the 2D model, we take the optical conductivity of multilayer graphene to be $\sigma_{t}=N \times \sigma_{g}$, where interlayer interactions are neglected (this has been shown experimentally to be a rather accurate description of multilayer graphene conductivity in the visible range \cite{Nair:08}). Figure \ref{fig:Sensi} shows the obtained $\Delta S$ as functions of the number of graphene layers. 

When graphene is modeled through the 2D or the 3D anisotropic models (green and cyan lines in Fig. \ref{fig:Sensi}, respectively), the addition of graphene layers practically does not affect the sensitivity of the system based only on gold's SPR. However, when the 3D isotropic model is applied, the increase is rather large (dark blue line, Fig. \ref{fig:Sensi}). Also, by changing the graphene refractive index to the one used in \cite{Bhavsar:19,Wu:10,SAIFUR:17} ($n^g_{3}  = 3+1.1491i$), we get an even larger sensitivity increase (red line, Fig. \ref{fig:Sensi}), which agrees with that reported by Ref. \cite{Bhavsar:19}. We believe this sensitivity enhancement, obtained solely with the isotropic model, is related to the fact that in this case the graphene film would effectively reduce the absolute value of the isotropic dielectric constant of gold at the surface, which is known to increase the sensitivity to changes in the analyte's refractive index \cite{POCKRAND:78,Raether:88}. Indeed, if we artificially reduce gold’s dielectric constant in our model without graphene, we also obtain an increase in the system's sensitivity. We, thus, highlight that modelling graphene as two-dimensional surface or as an anisotropic thin film are more accurate alternatives, and are found to lead to no significant SPR sensor sensitivity increases.\\

\section{Conclusions}

In this paper we theoretically studied the influence of graphene modelling to the sensitivity obtained in a refractometer based on gold's SPR. When graphene is modeled as a surface (2D model) or an anisotropic thin film (anisotropic 3D model) the system's sensitivity is practically unaffected by the presence and number of layers of graphene. However, when graphene is modelled as an isotropic thin film (isotropic 3D model), a significant sensitivity increase is observed, specially when the refractive index used is the same as that by previous works\cite{Bhavsar:19,Wu:10,SAIFUR:17}. Treating graphene as an isotropic film is known to lead to a number of previously reported un-physical results. Therefore, we suggest that modelling graphene through the 2D or 3D anisotropic models are more accurate alternatives, which must be adopted for the study of graphene-assisted SPR sensors.

\section{Acknowledgements}
This work was funded by FAPESP (grant nos. 2018/07276-5 and 2018/25339-4), the Brazilian Nanocarbon Institute of Science and Technology (INCT/Nanocarbon), and CAPES-PrInt (grant no. 88887.310281/2018-00). NMRP acknowledges PORTUGAL 2020, FEDER, and the FCT through projects: UIDB/04650/2020 strategic project, QML-HEP - CERN/FIS-COM/0004/2021, PTDC/FIS-MAC/2045/2021, and POCI-01- 0145-FEDER-028114,  and the European Commission through the project GrapheneDriven Revolutions in ICT and Beyond (Ref. No. 881603, CORE 3).

\bibliographystyle{unsrt}
\bibliography{artigobib}

\end{document}